# A faster than "World fastest derivation of the Lorentz transformation"

## Bernhard Rothenstein,
"Politehnica" University Timisoara Romania

Macdonald[1] considers that his derivation of the Lorentz-Einstein transformation (LET) is the fastest one in the world. It is based on the assumptions:
A. The speed of light is the same in all inertial reference frames.
B. A clock moving with constant speed **V** in an inertial reference frame K runs at a constant rate $\gamma = \gamma(V)$ with respect to the synchronized clocks of K which it passes.
Assumption B is a cryptic way of presenting the time dilation effect.

The purpose of our note is to present a derivation of the (LET) based on the formula that accounts for the length contraction effect

$$L = L_0 \sqrt{1 - \frac{V^2}{c^2}} \qquad (1)$$

where $L$ and $L_0$ represent the length of the same rod measured by an observer relative to whom the rod moves with constant speed $V$ in its direction ($L$) and by an observer relative to whom it is at rest (proper length $L_0$), respectively.

The scenario we follow[2] involves two parallel rods **1** and **2** with proper lengths $L_{0,1}$ and $L_{0,2}$, respectively. (Figure 1) Rod **1** is at rest in the K(OX) frame and located along the OX axis with its left end at O. Rod **2** moves with constant speed $V$ in the positive direction of the OX axis. The clocks involved are $C_0(0)$ and $C(L_{0,1})$ located at the left and right end of rod **1**, respectively and $C'_0(0)$ and $C'(L_{0,2})$ located at the left and right end of rod **2**, respectively. When clocks $C_0(0)$ and $C'_0(0)$ both read $t = t' = 0$, the left ends of the two rods are located at the same point in space. At that very moment, a source of light $S(0)$ located at the left end of rod **1** and at rest relative to it emits a light signal (Figure 1a). When the right ends of the two rods are located at the same point in space, clock $C(L_{0,1})$ reads $t = L_{0,1} c^{-1}$, clock $C'(L_{0,2})$ reads $t' = L_{0,2} c^{-1}$ and rod **2** has advanced with

$$V L_{0,1} c^{-1}. \qquad (2)$$

The length of rod **2** measured by observers from the stationary reference frame is in accordance with (1)

$$L_2 = L_{0,1}\sqrt{1 - \frac{V^2}{c^2}}. \tag{3}$$

If the length of a rod is defined as a difference between the space coordinates of its ends, we have

$$L_{0,1} = x - 0 = ct \tag{4}$$

$$L_{0,2} = x' - 0 = ct'. \tag{5}$$

Figure 1b displays lengths measured by observers of the same reference frame, and so we can add them, yielding the result

$$L_{0,1} = Vtc^{-1} + L_{0,2}\sqrt{1 - \frac{V^2}{c^2}}. \tag{6}$$

Expressing (6) as a function of the space-time coordinates defined above, this equation becomes

$$x = Vt + x'\sqrt{1 - \frac{V^2}{c^2}} \tag{7}$$

or

$$x = \frac{Vt}{c} + ct'\sqrt{1 - \frac{V^2}{c^2}}. \tag{8}$$

From (7), we obtain

$$x' = \frac{x - Vt}{\sqrt{1 - \frac{V^2}{c^2}}} \tag{9}$$

whereas (8) leads to

$$t' = \frac{t - \frac{V}{c^2}t}{\sqrt{1 - \frac{V^2}{c^2}}} \quad . \tag{10}$$

We have derived the Lorentz-Einstein transformations for the space time coordinates of the same event $E(x,t)$ in K and $E'(x',t')$ in K' associated with the fact that the right ends of the two rods are located at the same point in space. The way in which we have derived these transformations clearly shows the conditions under which they hold and the part played by rods and clocks in the derivations.

If we measure the "length" of a derivation by the number of assumptions made throughout, we can consider that our derivation makes the single length contraction assumption.

By knowing the exact meaning of the physical quantities involved in the LET, we can use them to derive formulas that account for different relativistic effects, without obscuring the physics behind them. By considering the same scenario from the reference frame where rod **2** is at rest, we can derive the inverse LET.

References

[1] Alan Macdonald, "World fastest derivation of the Lorentz transformations," Am.J.Phys.**49,** 483 (1981)

[2] Leo Karlov, "Paul Kard and Lorentz free special relativity," Phys.Educ. **24,** 165-168 (1989)

*Figure 1. Scenario for deriving the Lorentz-Einstein transformations*

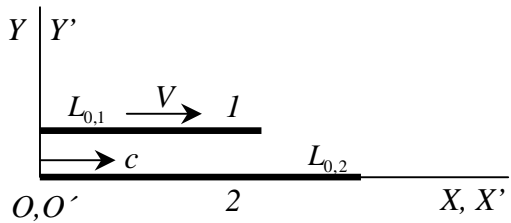